# Electrical spin injection and transport in Germanium

Yi Zhou[1a], Wei Han[2a], Li-Te Chang[1], Faxian Xiu[1], Minsheng Wang[1], Michael Oehme[3], Inga A. Fischer[3], Joerg Schulze[3], Roland. K. Kawakami[2], and Kang L. Wang[1*]

[1]Device Research Laboratory, Department of Electrical Engineering, University of California, Los Angeles, California, 90095, USA [2]Department of Physics and Astronomy, University of California, Riverside, California, 92521, USA [3]Institut fuer Halbleitertechnik (IHT), Universitaet Stuttgart, Stuttgart, 70569, Germany

We report the first experimental demonstration of electrical spin injection, transport and detection in bulk germanium (Ge). The non-local magnetoresistance in n-type Ge is observable up to 225K. Our results indicate that the spin relaxation rate in the n-type Ge is closely related to the momentum scattering rate, which is consistent with the predicted Elliot-Yafet spin relaxation mechanism for Ge. The bias dependence of the nonlocal magnetoresistance and the spin lifetime in n-type Ge is also investigated.

[a] these authors contributed equally to this work

* wang@ee.ucla.edu

Information processing based on the electron's spin degree of freedom is envisioned to offer a new paradigm of electronics beyond the conventional charge-based device technologies [1, 2]. To add spin functionality into semiconductor-based field effect transistors (spin-FET) [3-5] is considered as one of the approaches to overcome the ultimate scaling limits of the mainstream silicon (Si)-based complementary metal-oxide-semiconductor (CMOS) technology [6]. Electrical injection and transport of spin-polarized electrons from ferromagnetic metals (FMs) into the semiconductors is a prerequisite for developing such an approach [1, 2]. Although significant progress has been achieved in GaAs [7, 8] and Si [9-12], little progress has been made in germanium (Ge), despite its ultimate importance in the semiconductor industry owing to the high charge carrier mobilities and the compatibility with the established CMOS technology. In addition, Ge is also expected to have enhanced spin lifetime and transport length due to the weak spin-orbital interaction resulting from the lattice inversion symmetry [13]. Liu *et al*. [14] reported the spin diffusion length in Ge nanowires (NW) based on local magnetoresistance (MR) between two Co/MgO/Ge NW contacts. However, unambiguous evidence of spin injection and transport based on non-local Hanle precession has not been reported. Furthermore, the spin lifetime, as well as the underlying physics governing the spin relaxation in Ge, remains to be explored. This information is ultimately important to a successful realization of Ge-based Spintronic devices.

In this paper, we report the first demonstration of electrical spin injection to bulk

Ge by using an epitaxially grown Fe/MgO/n-Ge tunnel junction. The spin dependent properties of Ge are characterized by the non-local spin transport measurements. The non-local MR in Ge is observed up to 225K. Both the non-local MR and the spin lifetime are found to be weakly dependent on temperature at low temperature region (T<30K). However, the dependence becomes much stronger as the temperature increases. This is attributed to the dominance of spin relaxation by ionized impurity scattering at low temperatures and phonon scattering at higher temperatures. Our results show a close relation between the spin relaxation rate and the momentum relaxation rate, which is consistent with the predicted Elliot-Yafet spin relaxation mechanism for Ge [15, 16]. We also examine the bias dependence of the non-local MR and spin lifetime. The smaller non-local MR and shorter spin lifetime under forward biases are caused by the fast spin relaxation rate in the highly doped Ge surface layer.

An unintentionally doped *n*-type Ge wafer is used as the starting substrate. A lightly doped $n^-$ ($n=1\times10^{16} cm^{-3}$) Ge layer (300 nm) is grown on this substrate as the spin transport channel. Above this layer is a transition layer (15 nm) to a degenerately doped $n^+$ ($n=2\times10^{19} cm^{-3}$) surface layer (15 nm). All these layers are grown by low temperature solid source molecular beam epitaxy [17]. Two devices (A and B) are fabricated on this wafer with the same processes. First, a device channel is defined by photolithography and etched by reactive ion etching. The width of the channel is 5 μm and 15 μm for device A and B, respectively. The height of the channel mesa is 60 nm for both devices. Four electrodes are then fabricated on the channel by the standard e-beam lithography and liftoff process. The outer two electrodes are made of Au/Ti. The center two spin-dependent electrodes are made of MgO (1nm) and Fe (100 nm), which are deposited in a molecular beam epitaxy system, and capped by 5 nm $Al_2O_3$. The as-grown MgO is single crystalline and possessing 45 degree in-plane rotation of the unit cell with respect to that of the Ge [18]. A schematic of the atomic configuration is shown in Figure 1a. This high quality Fe/MgO/Ge junction not only alleviates the Fermi level pinning at the Ge surface to favor electronic transport [19], but also leads to an enhanced spin polarization of the injected electrons due to the symmetry induced spin filtering [20]. To characterize the spin injection and transport in Ge, we employ the non-local measurement technique [7, 21-24]. The center-to-center distance ($L$) between the spin injector (E2) and spin detector (E3) is 420 nm and 1 μm for device A and device B, respectively. A schematic diagram of the device structure and measurement scheme is shown in Figure 1(b). The standard low-frequency lock-in technique is used for the measurement.

Figure 1(c) shows a scanning electron microscope (SEM) image of the device A. The widths of the spin injector (E2) and spin detector (E3) are 400 nm and 250 nm, respectively. Temperature dependent *I-V* characteristics measured between the spin injector (E2) and E1 are shown in Figure 1(d). Since E1 is made of Au/Ti and the size is much larger than E2, we consider the *I-V* characteristics are dominated by the contact resistance of the spin injector (E2). The nonlinearity and weak temperature dependence of the I-V characteristics confirm the tunneling nature of this contact [25], which is necessary to overcome the conductivity mismatch problem for spin injection from FMs to semiconductors [26-28].

To characterize spin injection and transport in Ge, we first perform the non-local spin valve measurement. In this measurement, a charge current is applied between the spin injector (E2) and E1 (as shown in Figure 1b), resulting in a spin accumulation in the Ge at E2 by means of spin injection (E2 under a reverse bias) or spin extraction (E2 under a forward bias) [29, 30]. In either case, once spin accumulation is created, the spin-polarized electrons start to diffuse isotropically in the Ge channel. The spin detector (E3) is placed outside the charge current path, and it detects a voltage potential which is proportional to the projection of the spin accumulation in the Ge onto its magnetization direction. Therefore, if the spin accumulation of the injected electrons is sizeable when they diffuse to E3, a bipolar non-local voltage $V_{NL}$ should be observed which changes sign when the magnetization directions of the spin injector (E2) and detector (E3) switch from parallel to anti-parallel. To modulate the magnetization directions of the spin injector (E2) and detector (E3), an external magnetic field ($B_y$) along the easy axis of the electrodes (y direction as indicated in Figure 1b) is swept and the $V_{NL}$ is recorded as a function of $B_y$. Figure 2(a) shows the non-local spin valve signal measured on device A at 4K, with a reverse DC bias current ($I_{DC}$) of -20 µA and AC current ($I_{AC}$) of 10 µA. The non-local resistance $R_{NL}$ is defined as the $V_{NL}$ divided by $I_{AC}$. The difference of $R_{NL}$ between the parallel and antiparallel configuration is defined as the non-local MR ($\Delta R_{NL}$) and measured to be 0.94 Ω in this case.

Figure 2 (b) shows the non-local spin valve signals measured on device A at different temperatures. The signal is observable up to 225K. Figure 2 (c) summarizes the $\Delta R_{NL}$ as a function of the temperature for device A. The $\Delta R_{NL}$ is weakly dependent on temperature at low temperature region, which increases slightly from 0.83 Ω at 1.5K to 1.04 Ω at 10K, and then decreases to 0.72 Ω at 30K. However, as the temperature further increases, the $\Delta R_{NL}$ drops abruptly and is not observable for T>225 K. Similar temperature dependence of $\Delta R_{NL}$ is also observed in device B with a longer transport channel ($L = 1$ μm, as shown in Figure 2d).

To further explore the spin transport properties in Ge, we perform the non-local Hanle measurement. In this measurement, a small transverse (in $z$ direction as shown in Figure 1b) magnetic field ($B_z$) is applied to induce the precession of the injected spin by the Hanle effect [23, 31]. The precession and dephasing of the spins during their transport in Ge is manifested as the magnetic field ($B_z$) dependence of the $V_{NL}$ (or $R_{NL}$, equivalently). Figure 3(a) shows the Hanle precession curves of device B at 4K under a reverse bias of -130 μA, which provide the unambiguous evidence of spin injection and transport in Ge. The red and black symbols are for signals taken when the injector/detector magnetizations are in parallel and anti-parallel configurations, respectively. A spin lifetime ($\tau_s$) of 1.08 ns is extracted by fitting the Hanle curves based on the 1-D spin drift diffusion model [7, 21, 24], under which

$$R_{NL} \propto \pm \int_0^\infty \frac{1}{\sqrt{4\pi Dt}} \exp[-\frac{L^2}{4Dt}] \cos(\omega_L t) \times \exp(-\frac{t}{\tau_s}) dt \qquad (1)$$

In the above equation, + (-) sign is for the parallel (antiparallel) magnetization configuration, $D$ is the diffusion constant, $\omega_L = g\mu_B B_z / \hbar$ is the Larmor frequency (where $g$=1.6 is the Landé g-factor for Ge [32], $\mu_B$ is the Bohr magneton and $\hbar$ is the reduced Planck constant). The temperature dependent spin lifetimes for device A and device B (obtained under reverse biases) are shown in Figure 3(b) in solid circles and open squares, respectively. Similar to the temperature dependence of $\Delta R_{NL}$, the dependence of the spin lifetime on temperature is rather weak at low temperatures, while it becomes much stronger as the temperature increases. This can be explained as in the following. For Ge, which possesses the lattice inversion symmetry, the spin relaxation is predicted to be dominated by the Elliot-Yafet mechanism [15, 16], under which the spin relaxation rate ($1/\tau_s$) is proportional to the momentum relaxation rate. The two major sources of momentum relaxation are the ionized impurity scattering and the phonon scattering. And the temperature dependence of the ionized impurity scattering rate is found to be much weaker than that of the phonon scattering in n-type Ge [33]. It is expected that at low temperature region, ionized impurity is the dominant scattering source, therefore a weak temperature dependence of the spin relaxation rate (or spin lifetime, equivalently) is observed. As the temperature increases, phonon scattering becomes dominant, resulting in a much higher temperature dependence of the spin lifetime. Our results are consistent with the predicted Elliot-Yafet spin relaxation mechanism for Ge.

Finally we study the bias dependence of the $\Delta R_{NL}$ and the spin lifetime. Figure 4(a) shows the DC bias dependent $\Delta R_{NL}$ of device A at different temperatures. Since we use the lock-in technique, the measured $V_{NL}$ is characteristic of the slope of $V_{NL}$ versus $I_{DC}$ curve from the DC measurement. The inset of Figure 4(a) shows the restored DC relation between the $V_{NL}$ and $I_{DC}$ at 10K by numerically integrating our $V_{NL}$ over $I_{DC}$. The bias dependence of $V_{NL}$ at reverse bias is consistent with the reported results on the Fe/GaAs system [7]. However, our data do not display the nonmonotonic behavior at forward biases, which was attributed to the localized electrons in the surface bands due to the doping profile in the Fe/GaAs system [34]. It is noted that the $\Delta R_{NL}$ is much smaller at forward biases as compared to those at reverse biases. Figure 4(b) and (c) shows the Hanle precession curves at 10K with a DC reverse bias of -20 µA and a forward bias of +20 µA, respectively. It is also found that the spin lifetime extracted from forward bias (332 ps) is shorter than that from the reverse bias (773 ps). The bias dependence of the $\Delta R_{NL}$ and spin lifetime can be explained by the doping dependent spin relaxation as in the following. When a reverse bias is applied, the depletion region in the Ge extends and spin polarized electrons are injected into the lightly doped Ge channel (inset of Figure 4b). However, when a forward bias is applied, the depletion region is reduced and the spin accumulates mainly at the highly doped surface layer (inset of Figure 4c), where a faster spin relaxation rate is expected due to the larger momentum scattering by ionized impurities.

In conclusion, we have successfully achieved electrical spin injection, transport and detection in bulk n-type Ge by using an Fe/MgO/n-Ge tunnel junction. Investigating the temperature and bias dependence of the non-local spin valve signals and the spin lifetimes, we show that the spin relaxation in Ge is consistent with the predicted Elliot-Yafet mechanism. Our results present a major step towards achieving Ge-based Spintronics devices for a new paradigm of non-volatile electronics beyond

CMOS technology.

**Acknowledgements:** We gratefully acknowledge the financial support from the Western Institution of Nanoelectronics (WIN) through NRI. WH and RKK acknowledge support from NSF (CAREER DMR-0450037). The technical support from Jens Werner (IHT) and Dr. Olaf Kirfel (IHT) is also acknowledged. We thank Dr. Dmitri Nikonov, Dr. Ajey Jacob, and Dr. Charles Kuo of Intel Corporation and Dr. An Chen of Globalfoundries Corporation for the valuable discussion.

**References:**

[1] S. A. Wolf *et al*., Science. **294**, 1488 (2001).

[2] I. Zutic, J. Fabian, and S. D. Sarma, Rev. Mod. Phys. **76**, 323 (2004).

[3] S. Datta and B. Das, Appl. Phys. Lett. **56**, 665 (1990).

[4] S. Sugahara and M. Tanaka, Appl. Phys. Lett. **84**, 2307 (2004).

[5] K. C. Hall and M. E. Flatte, Appl. Phys. Lett. **88**, 162503 (2006).

[6] International Technology Roadmap for Semiconductors, 2009 Edition,

http://www.itrs.net/reports.html. See Emerging Research Devices.

[7] X. Lou *et al*., Nature Phys. **3**, 197 (2007).

[8] X. Lou *et al*., Phys. Rev. Lett. **96**, 176603 (2006).

[9] I. Applebaum, B. Huang, and D. J. Monsma, Nature. **447**, 295 (2007).

[10] B. T. Jonker *et al*., Nature Phys. **3**, 542 (2007).

[11] S. P. Dash *et al*., Nature. **462**, 491 (2009).

[12] T. Suzuki *et al*., Appl. Phys. Express. **4**, 023002 (2011).

[13] M. I. Dyakonov, *Spin Physics in Semiconductors* (Springer-Verlag, Berlin, 2008).

[14] E. -S. Liu *et al*., Nano Lett. **10**, 3297 (2010).

[15] R. J. Elliott, Phy. Rev. **96**, 266-279 (1954).

[16] Y. Yafet, in *Solid State Physics*, edited by F. Seitz. And D. Turnbull. (Academic,

New York, 1963), Vol. 14, p. 1.

[17] M. Oehme, J. Werner, and E. Kasper, J. Crys. Growth. **310**, 4531-4534 (2008).

[18] W. Han *et al*., J. Crys. Growth. **312**, 44-47 (2009).

[19] Y. Zhou *et al*., Appl. Phys. Lett. **96**, 102103 (2010).

[20] W. H. Butler *et al.*, Phys. Rev. B. **63**, 054416 (2001).

[21] F. J. Jedema *et al.*, Nature. **416**, 713 (2002).

[22] O. M. J. van't Erve *et al.*, IEEE. Trans. Electron. Device. **56**, 2343 (2009).

[23] N. Tombros *et al.*, Nature. **448**, 571 (2007).

[24] W. Han *et al.*, Phys. Rev. Lett. **105**, 167202 (2010).

[25] B. J. Jonsson-Akerman *et al.*, Appl. Phys. Lett. **77**, 1870 (2000).

[26] G. Schmidt *et al.*, Phys. Rev. B. **62**, R4790-4793 (2000).

[27] E. I. Rashba, Phys. Rev. B. **62**, R16267 (2000).

[28] A. Fert and H. Jaffres, Phys. Rev. B. **64**, 184420 (2001).

[29] C. Ciuti, J. P. McGuire, and L. J. Sham, Phys. Rev. Lett. **89**, 156601 (2002).

[30] J. Stephens *et al.*, Phys. Rev. Lett. **93**, 097602 (2004).

[31] M. Johnson, and R. H. Silsbee, Phys. Rev. Lett. **55**, 1790-1793 (1985).

[32] G. Feher, D. K. Wilson, and E. A. Gere, Phys. Rev. Lett. **3**, 25 (1959).

[33] P. P. Debye and E. M. Conwell, Phys. Rev. **93**, 693-706 (1954).

[34] H. Dery and L. J. Sham, Phys. Rev. Lett. **98**, 046602 (2007).

**Figure Captions**

**Figure 1 (color online).** (a) Schematic atomic configuration of the Fe/MgO/Ge junction, showing 45 degree rotation of the MgO unit cell with respect to that of the Ge. (b) A schematic diagram of the device structure and the non-local measurement scheme. The center-to-center distances between the spin injector and detector are 420 nm and 1 μm for device A and B, respectively. (c) A SEM image of device A. The widths of the injector (E2) and detector (E3) are 400 nm and 250 nm, respectively. (d) Temperature dependent I-V curves measured between spin injector (E2) and E1.

**Figure 2 (color online).** (a) Non-local spin valve signal measured on device A at 4K with a DC injection current of -20 μA and AC injection current of -10 μA. The blue arrows indicate the magnetization directions of the injector and detector. (b) Temperature dependent non-local spin valve signals on device A. The curves are offset for clarity. (c) and (d), Temperature dependent non-local magnetoresistance ($\varDelta R_{NL}$) of device A and B, respectively.

**Figure 3 (color online).** (a) Non-local Hanle precession curves measured on device B at 4K with a DC injection current of -130 μA. The red and black symbols are for signals measured when the injector and detector are in parallel and antiparallel configurations, respectively. The solid lines are fitting based on the 1-D spin drift diffusion model, from which the spin lifetime is extracted to be 1.08 ns. (b) Temperature dependent spin lifetimes measured on device A (solid circles) and B (open squares), respectively.

**Figure 4 (color online).** (a) The DC bias dependent $\varDelta R_{NL}$ of device A at different temperatures. The inset shows the restored DC relation between the $V_{NL}$ and $I_{DC}$ at 10K by numerically integrating our $V_{NL}$ over $I_{DC}$. (b) and (c), Hanle precession curves measured on device A at 10K under a reverse bias of -20 µA and a forward bias of +20 µA, respectively. The spin lifetimes extracted from the fittings (solid lines) based on the 1-D spin drift diffusion model are 773 ps and 332 ps for reverse and forward biases, respectively. The insets show the locations of the spin accumulation.

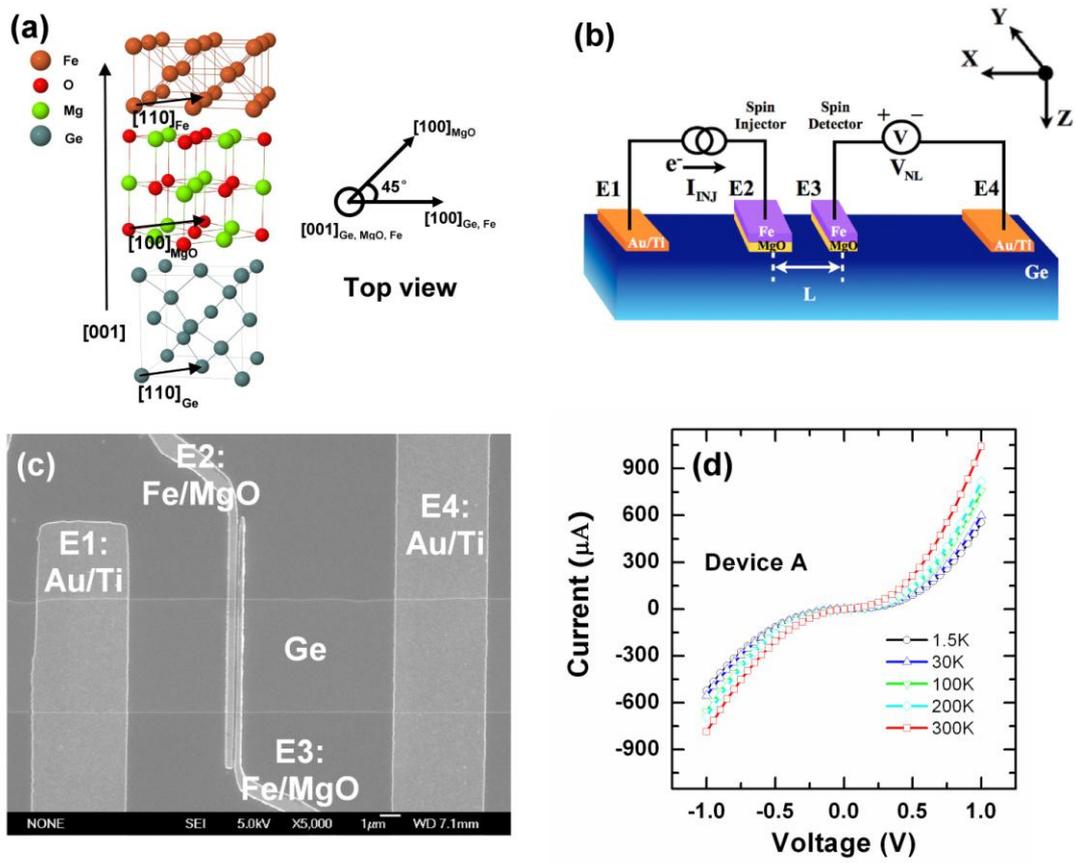

**Figure 1**

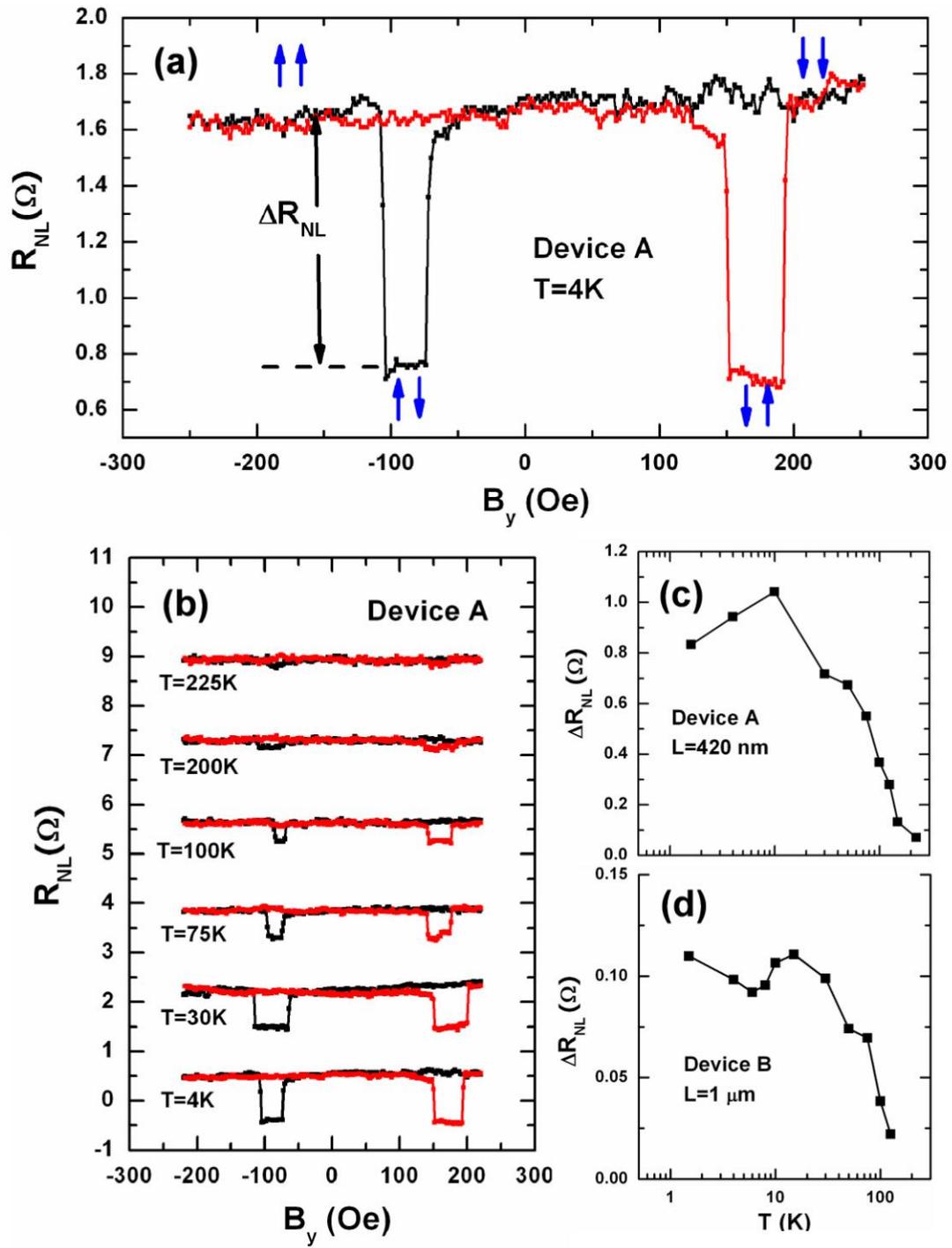

**Figure 2**

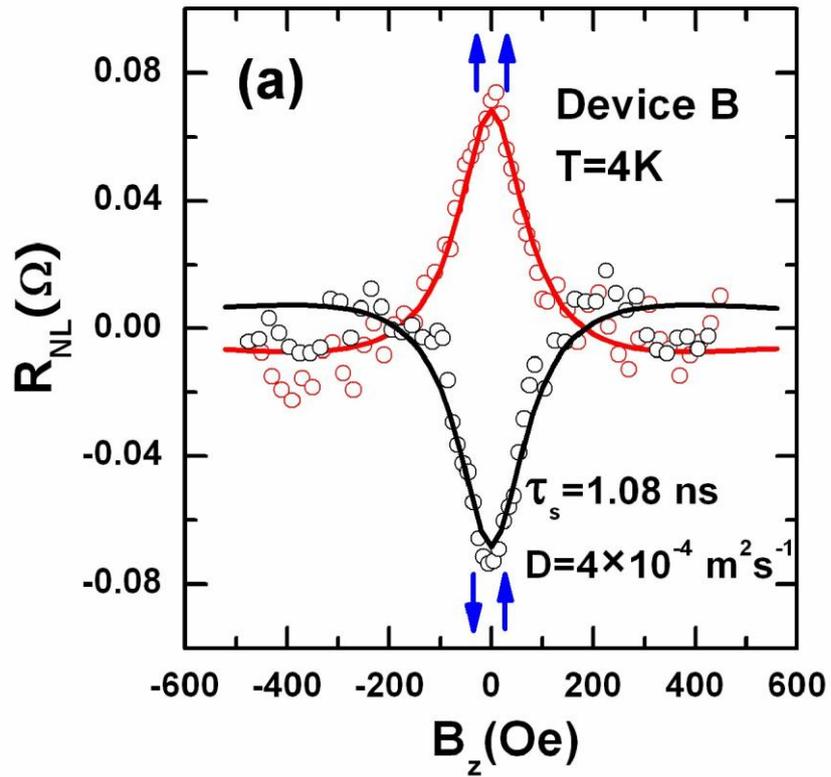
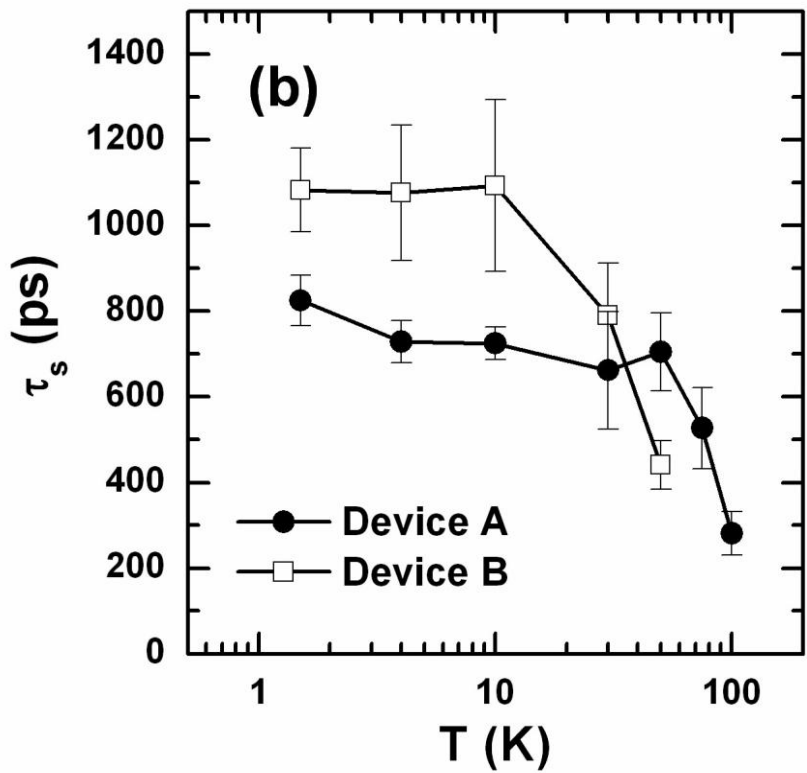

**Figure 3**

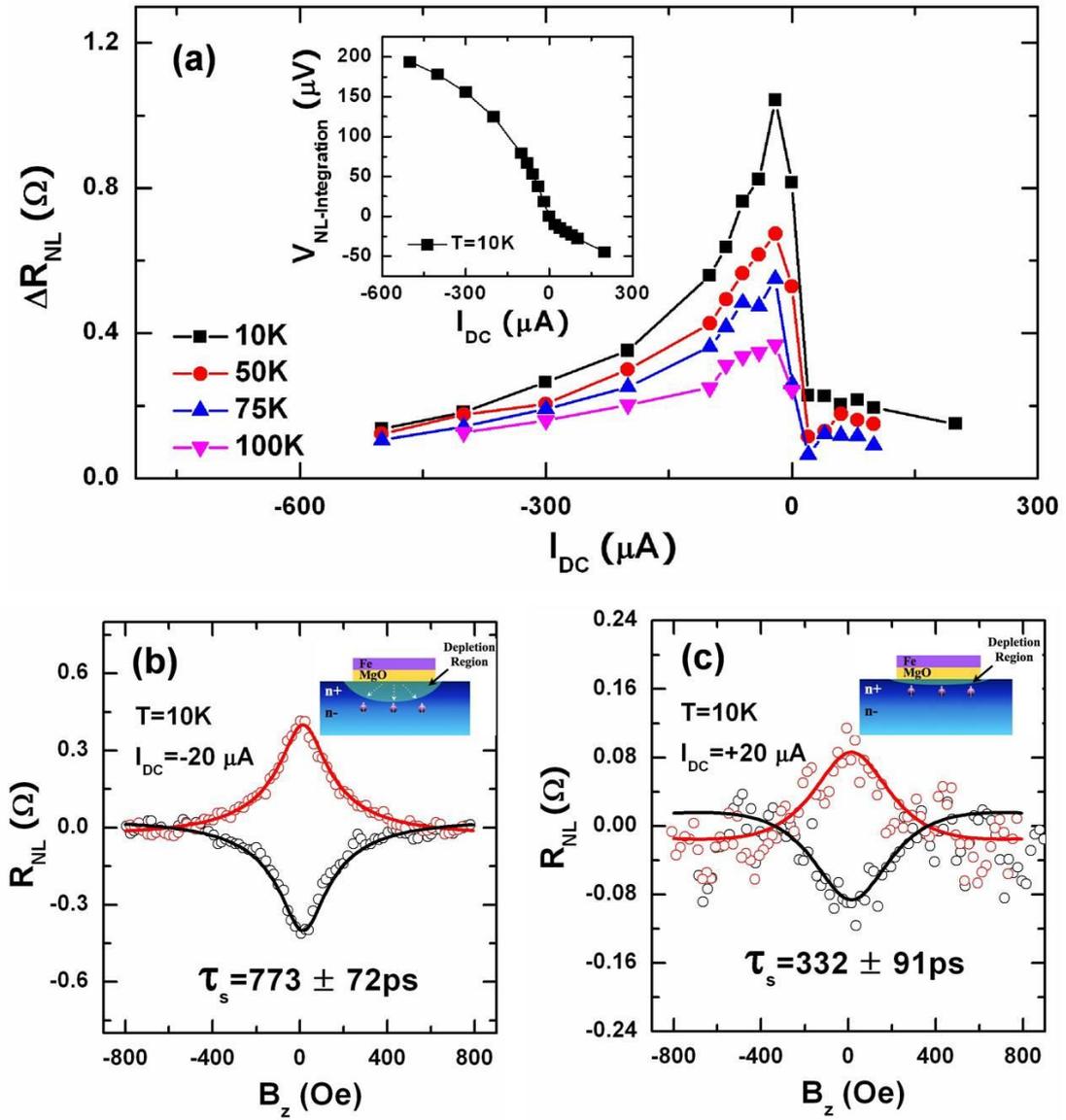

**Figure 4**